\documentclass[twocolumn]{aastex62}

\usepackage{graphicx,latexsym,amssymb,epsfig}

\begin{document}
\title{Effects of symmetry energy on equation of state for simulations of core-collapse
supernovae and neutron-star mergers}

\author[0000-0003-2717-9939]{Hong Shen}
\affiliation{School of Physics, Nankai University, Tianjin 300071, China; shennankai@gmail.com}

\author[0000-0001-6044-252X]{Fan Ji}
\affiliation{School of Physics, Nankai University, Tianjin 300071, China}

\author[0000-0002-1709-0159]{Jinniu Hu}
\affiliation{School of Physics, Nankai University, Tianjin 300071, China; hujinniu@nankai.edu.cn}

\author[0000-0002-9224-9449]{Kohsuke Sumiyoshi}
\affiliation{National Institute of Technology, Numazu College, Shizuoka 410-8501, Japan}

%%%%%%%%%%%%%%%%%%%%%%%%%%%%%%%%%%%%%%%%%%%%%%%%%%%%%%%%%%%%%%%%%%%%%%%%%%%%%%%%
\begin{abstract}
We construct a new equation of state (EOS) for numerical simulations of core-collapse
supernovae and neutron-star mergers
based on an extended relativistic mean-field model with a small symmetry energy
slope $L$, which is compatible with both experimental nuclear data and recent
observations of neutron stars. The new EOS table (EOS4) based on the extended
TM1 (TM1e) model with $L=40$ MeV is designed in the same tabular form and compared with
the commonly used Shen EOS (EOS2) based on the original TM1 model with $L=110.8$ MeV.
This is convenient and useful
for performing numerical simulations and examining the influences of symmetry
energy and its density dependence on astrophysical phenomena. In comparison with
the TM1 model used in EOS2, the TM1e model provides a similar maximum neutron-star
mass but smaller radius and tidal deformability for a $1.4 M_\odot$ neutron
star, which is more consistent with current constraints.
By comparing the phase diagram and thermodynamic quantities
between EOS4 and EOS2, it is found that the TM1e model predicts relatively larger
region of nonuniform matter and softer EOS for neutron-rich matter.
Significant differences between EOS4 and EOS2 are observed in the case with
low proton fraction, while the properties of symmetric matter remain unchanged.
\end{abstract}

\keywords{equation of state --- supernovae: general --- stars: neutron}

%%%%%%%%%%%%%%%%%%%%%%%%%%%%%%%%%%%%%%%%%%%%%%%%%%%%%%%%%%%%%%%%%%%%%%%%%%%%%%%%
\section{Introduction}
\label{sec:1}

The equation of state (EOS) is a critical input for astrophysical simulations
such as core-collapse supernovae and neutron-star mergers, which require information
over wide ranges of temperature, proton fraction, and baryon density.
The EOS should include reasonable descriptions for both nonuniform matter at subsaturation
densities and uniform matter at high densities.
Due to the complex phase structure of stellar matter, it is not an easy task to construct
the EOS covering the full range of thermodynamic conditions.
Currently, there are a set of EOSs available for supernova
simulations, which have been summarized in the review by Oertel et al.~\citep{oert17}.
One of the most commonly used EOSs is that of Lattimer and Swesty~\citep{latt91},
which was based on the compressible liquid-drop model with a Skyrme force.
Another commonly used EOS is often referred to as Shen EOS~\citep{shen98a,shen98b,shen11},
which used a relativistic mean-field (RMF) model and Thomas-Fermi approximation
with a parameterized nucleon distribution for the description of nonuniform matter.
Both EOSs employed the so-called single nucleus approximation (SNA), where only a
single representative nucleus was considered instead of an ensemble of nuclei.
It was shown that SNA could adequately describe the thermodynamics of the
system~\citep{burr84}.
Recently, EOS tables were developed beyond the SNA by including multiple
nuclei in nuclear statistical equilibrium (NSE) based on some RMF or
Skyrme parameterizations~\citep{hemp10,furu11,furu13,furu17a,stei13,schn17}.
In~\citet{shenG11a,shenG11b}, the authors employed a hybrid approach for constructing
EOS tables, where NSE was used at low densities and SNA was adopted at intermediate
densities near the transition to uniform matter.
It is known that considering detailed nuclear composition plays an important role
in determining the electron-capture rates and neutrino-matter
interactions, but it has less influence on thermodynamic quantities of dense matter.
In addition, microscopic approaches based on realistic nuclear forces have been also
applied to construct the EOS tables for astrophysical simulations~\citep{toga17,furu17b}.
In~\citet{schn19}, the authors developed the EOS tables based on the Skyrme-type
parameterization of the nuclear force, where the parameters were tuned to reproduce
the Akmal, Pandharipande, and Ravenhall (APR) EOS.

The recent developments in astrophysical observations provide quantitative
constraints on the EOS of dense matter.
One strong constraint comes from the mass measurements of massive
pulsars, PSR J1614-2230~\citep{demo10,fons16}, PSR J0348+0432~\citep{anto13},
and PSR J0740+6620~\citep{crom19},
which requires the maximum neutron-star mass to be larger than $\sim 2 M_\odot$.
Another constraint is provided by the radius estimations from quiescent low-mass
X-ray binaries and objects with photospheric radius expansion bursts,
which suggest small neutron-star radii, but it has much larger uncertainties
than the mass measurements~\citep{fort15}.
Furthermore, the first detection of gravitational waves from a binary neutron-star
merger, known as GW170817, provides an upper limit on the tidal deformability
of neutron stars~\citep{abbo17,abbo18}, which implies small neutron-star
radii also~\citep{fatt18,most18}.
More recently, the second detection of gravitational waves, GW190425, was
reported by LIGO and Virgo Collaborations~\citep{abbo19}, which implies a rather
large total mass of the binary system of $3.4^{+0.3}_{-0.1} M_\odot$
and may offer valuable information for the EOS at high densities.
The recent observations by {\it Neutron Star Interior Composition Explorer} ({\it NICER})
for PSR J0030+0451 provided a simultaneous measurement of the mass and radius of
a neutron star. From independent analyses of the {\it NICER} data on PSR J0030+0451,
Riley et al.~\citep{Rile19} reported a mass of $1.34^{+0.15}_{-0.16}  M_\odot$
with an equatorial radius of $12.71^{+1.14}_{-1.19}$ km,
while Miller et al.~\citep{Mill19} reported a mass of
$1.44^{+0.15}_{-0.14}  M_\odot$ with a radius of $13.02^{+1.24}_{-1.06}$ km.
It is interesting to notice that constraints on the neutron-star radii
from various observations are consistent with each other.

At present, some available EOS tables for supernova simulations are inconsistent
with these constraints. The EOS based on the FSU parametrization predicts
a maximum neutron-star mass of only $1.75 M_\odot$, which was improved by introducing
an additional phenomenological pressure at high densities~\citep{shenG11b}.
The RMF parametrizations, NL3 and TM1, lead to too large neutron-star radii
in comparison with the extracted values from astrophysical observations~\citep{oert17}.
In our previous work~\citep{shen11}, the EOS tables (EOS2 and EOS3)
were constructed by employing the TM1 model, while the nonuniform matter was described
in the Thomas--Fermi approximation with a parameterized nucleon distribution.
In EOS2, only nucleonic degrees of freedom were taken into account, while additional
contributions from $\Lambda$ hyperons were included at high densities in EOS3.
The TM1 model can provide a satisfactory description for finite nuclei
and a maximum neutron-star mass of $2.18 M_\odot$ with nucleonic degrees of freedom
only, but the resulting neutron-star radii seem to be too large~\citep{suga94,shen98a}.
Therefore, we would like to improve our EOS table in order to be consistent
with all available constraints from astrophysical observations.

It is well known that the neutron-star radius is closely related to the density
dependence of nuclear symmetry energy~\citep{horo01}.
There exists a positive correlation between the slope parameter $L$ of
the symmetry energy and the neutron-star radius~\citep{alam16}.
Since the TM1 model has a rather large slope parameter $L=110.8$ MeV, it predicts
too large radii for neutron stars as compared to the estimations from astrophysical
observations. In the present work, we prefer to employ an extended version of the
TM1 model with $L=40$ MeV (hereafter referred to as the TM1e model),
where an additional $\omega$-$\rho$ coupling term is introduced to modify the
density dependence of the symmetry energy~\citep{bao14b}.
By adjusting simultaneously two parameters associated to the $\rho$ meson in the TM1e
model, we achieve the slope parameter $L=40$ MeV at saturation density and the same
symmetry energy as the original TM1 model at a density of 0.11 fm$^{-3}$.
It is noteworthy that the TM1e and original TM1 models have the same isoscalar
properties and fixed symmetry energy at $0.11\, \rm{fm}^{-3}$, so that
both models can provide very similar descriptions of stable nuclei.
There are also other extended TM1 models for varying the symmetry energy slope $L$ by
including $\omega$-$\rho$ or $\sigma$-$\rho$ coupling term~\citep{prov13,prov16},
where the coupling constants associated to the $\rho$ meson
are adjusted to yield the same symmetry energy
as the original TM1 model at a density of 0.1 fm$^{-3}$.
In our TM1e model, we prefer to fix the symmetry energy at the density of
$0.11\, \rm{fm}^{-3}$, since this choice can provide almost unchanged
binding energy of $^{208}$Pb for different $L$ (see Figure 1 of~\citet{bao14b}).
Furthermore, the TM1e model predicts much smaller neutron-star radii than the original
TM1 model due to the difference in the slope parameter $L$.
It is found that the TM1e model yields a radius of 13.1 km for a canonical $1.4 M_\odot$
neutron star, while the corresponding value of the original TM1 model is as large
as 14.2 km~\citep{ji19}. According to the constraints based on astrophysical
observations and terrestrial nuclear experiments~\citep{oert17,tews17,tami17},
the slope parameter $L=40$ MeV of the TM1e model is more favored than $L=110.8$ MeV
of the original TM1 model. Moreover, the neutron-star radius in the TM1e model
is well within the new observational data by {\it NICER}.

We have two aims in this article. The first is to construct a new EOS
table (hereafter referred to as EOS4) for numerical simulations of
core-collapse supernovae and neutron-star mergers
based on the TM1e model, which is compatible
with both experimental nuclear data and recent observations of neutron stars.
The second is to make a detailed comparison between the new EOS4
and previous EOS2 in~\citet{shen11}, so that we can examine the influences of
symmetry energy and its slope on various aspects of the EOS for
astrophysical simulations.
We emphasize that both EOS4 and EOS2 are constructed using the same treatment for
nonuniform matter and uniform matter with nucleonic degrees of freedom,
but employ different RMF models for nuclear interaction.
Since the TM1e and TM1 models have the same properties of symmetric nuclear matter
but different behavior of symmetry energy, the differences between these two EOS
tables are solely due to different density dependence of symmetry energy.
For convenience in use and comparison, the new EOS4 is designed in the same tabular
form covering the full range of temperature, proton fraction, and baryon density
as described in~\citet{shen11}. For simplicity, only nucleonic degrees of freedom
are taken into account in EOS4, while the appearance of hyperons and/or quarks
at high densities is neglected.
By applying the new EOS4 together with EOS2 in astrophysical simulations,
it is possible to estimate the effects of symmetry energy and its density dependence
on core-collapse supernovae, black hole formation, and binary neutron-star merger.

This paper is arranged as follows. In Section~\ref{sec:2}, we briefly
describe the framework for building the EOS table.
In Section~\ref{sec:3}, we discuss and compare the new EOS4
with previous EOS2 by examining the phase diagram, compositions,
and thermodynamic quantities.
Section~\ref{sec:4} is devoted to a summary.

%%%%%%%%%%%%%%%%%%%%%%%%%%%%%%%%%%%%%%%%%%%%%%%%%%%%%%%%%%%%%%%%%%%%%%%%%%%%%%%%
\section{Formalism}
\label{sec:2}

For making the article self-contained, we give a brief description of
the RMF model and Thomas--Fermi approximation used for constructing the EOS table.

\subsection{RMF model}
\label{sec:2.1}

We employ the RMF model with an extended TM1 parametrization, namely the TM1e model,
to describe the nuclear system, where nucleons interact through
the exchange of various mesons including the isoscalar-scalar meson $\sigma$,
isoscalar-vector meson $\omega$, and isovector-vector meson $\rho$~\citep{bao14a,bao14b}.
The nucleonic Lagrangian density reads
\begin{eqnarray}
\label{eq:LRMF}
\mathcal{L}_{\rm{RMF}} & = & \sum_{i=p,n}\bar{\psi}_i
\left[ i\gamma_{\mu}\partial^{\mu}-\left(M+g_{\sigma}\sigma\right) \right. \nonumber \\
&& \left. -\gamma_{\mu} \left(g_{\omega}\omega^{\mu} +\frac{g_{\rho}}{2}
\tau_a\rho^{a\mu}\right)\right]\psi_i  \nonumber \\
&& +\frac{1}{2}\partial_{\mu}\sigma\partial^{\mu}\sigma
-\frac{1}{2}m^2_{\sigma}\sigma^2-\frac{1}{3}g_{2}\sigma^{3} -\frac{1}{4}g_{3}\sigma^{4}
\nonumber \\
&& -\frac{1}{4}W_{\mu\nu}W^{\mu\nu} +\frac{1}{2}m^2_{\omega}\omega_{\mu}\omega^{\mu}
+\frac{1}{4}c_{3}\left(\omega_{\mu}\omega^{\mu}\right)^2  \nonumber
\\
&& -\frac{1}{4}R^a_{\mu\nu}R^{a\mu\nu} +\frac{1}{2}m^2_{\rho}\rho^a_{\mu}\rho^{a\mu} \nonumber \\
&& +\Lambda_{\rm{v}} \left(g_{\omega}^2 \omega_{\mu}\omega^{\mu}\right)
\left(g_{\rho}^2\rho^a_{\mu}\rho^{a\mu}\right),
\end{eqnarray}
where $W^{\mu\nu}$ and $R^{a\mu\nu}$ denote the antisymmetric field
tensors for $\omega^{\mu}$ and $\rho^{a\mu}$, respectively\footnote{
Note that the coupling constant for isovector-vector meson,
$g_{\rho}$, is different by a factor of 2 from the one in~\citet{shen11}.
We follow here the convention of~\citet{bao14a}. }.
Under the mean-field approximation, the meson fields are treated as classical
fields and the field operators are replaced by their expectation values.
In a static uniform system, the nonzero components are
$\sigma =\left\langle \sigma \right\rangle$, $\omega =\left\langle
\omega^{0}\right\rangle$, and $\rho =\left\langle \rho^{30} \right\rangle$.
We derive the equations of motion for mesons and the Dirac equation
for nucleons, which are coupled with each other and could be solved self-consistently.

\begin{figure}[htbp]
\centering
\includegraphics[width=8.6 cm]{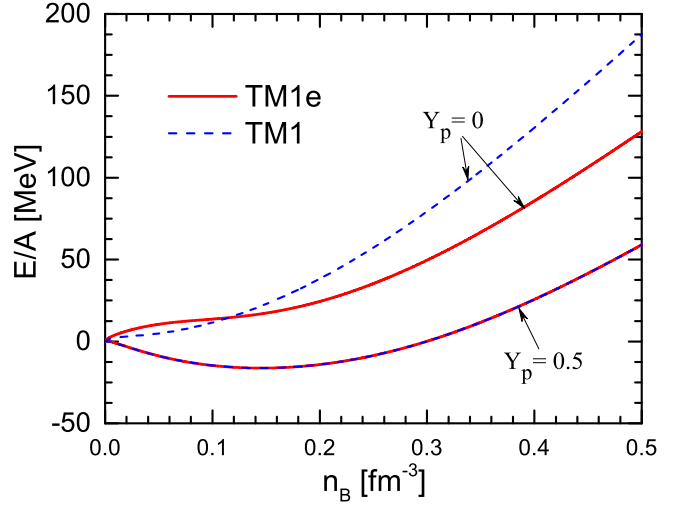}
\caption{Energy per baryon $E/A$ of symmetric nuclear matter and neutron matter
as a function of the baryon number density $n_B$ in the TM1e and TM1 models. }
\label{fig:1mat}
\end{figure}

\begin{figure}[htbp]
\centering
\includegraphics[width=8.6 cm]{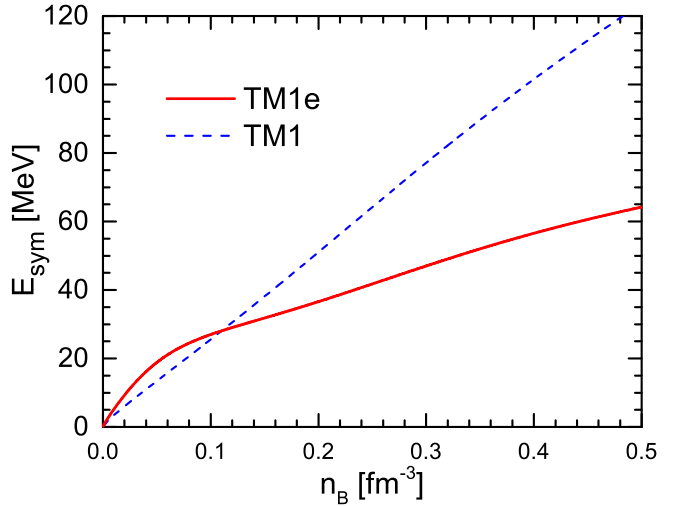}
\caption{Symmetry energy $E_{\rm{sym}}$ as a function of the baryon number density $n_{B}$
in the TM1e and TM1 models. }
\label{fig:2Esym}
\end{figure}

Compared with the original TM1 model adopted in~\citet{shen11},
an additional $\omega$-$\rho$ coupling term is introduced in the Lagrangian
density (\ref{eq:LRMF}), which plays a crucial role in determining the density
dependence of the symmetry energy~\citep{horo01,cava11,prov13,bao14a,bao14b}.
By adjusting the coupling constants, $g_{\rho}$ and $\Lambda_{\rm{v}}$,
it is possible to control the behavior of symmetry energy and its
density dependence. In the TM1e model, the slope parameter $L=40$ MeV and the
symmetry energy $E_{\text{sym}}=31.38$ MeV at saturation density are obtained,
which fall well within the constraints from various observations~\citep{oert17}.
The corresponding values in the original TM1 model are
$L=110.8$ MeV and $E_{\text{sym}}=36.89$ MeV, which are rather large and disfavored
by recent astrophysical observations.
In Table~\ref{tab:1}, we present the coupling constants of the TM1e and TM1 models
for completeness. It is shown that only $g_{\rho}$ and $\Lambda_{\rm{v}}$ related
to isovector parts are different, while all other parameters remain the same.
It is noteworthy that the TM1e model provides the same isoscalar saturation properties
and similar binding energies of finite nuclei as the original TM1 model,
whereas the density dependence of symmetry energy is very different.
In Figure~\ref{fig:1mat}, we plot the energy per baryon $E/A$ of symmetric nuclear
matter and neutron matter as a function of the baryon number density $n_B$.
It is shown that the behavior of symmetric nuclear matter is exactly the same between
the TM1e and TM1 models, while significant differences are observed in neutron matter.
This is related to different density dependence of symmetry energy between these
two models, which is displayed in Figure~\ref{fig:2Esym}.
One can see that the symmetry energy $E_{\text{sym}}$ in the TM1e model is slightly
larger at low densities and much smaller at high densities than that in the original
TM1 model. It is interesting and convenient to explore the influence of symmetry
energy and its density dependence on the properties of the EOS for supernova simulations
by using these two models.

For the Thomas--Fermi calculations of nonuniform matter, we need to input
the energy density and entropy density of uniform nuclear matter,
which are given in the TM1e model by
\begin{eqnarray}
\label{eq:ERMF}
\epsilon &=& \displaystyle{\sum_{i=p,n} \frac{1}{\pi^2}
  \int_0^{\infty} dk\,k^2\,
  \sqrt{k^2+{M^{\ast}}^2}\left( f_{i+}^{k}+f_{i-}^{k}\right) } \nonumber\\
 & &
  +\frac{1}{2}m_{\sigma}^2\sigma^2+\frac{1}{3}g_{2}\sigma^{3}
  +\frac{1}{4}g_{3}\sigma^{4} 
  +\frac{1}{2}m_{\omega}^2\omega^2+\frac{3}{4}c_{3}\omega^{4} \nonumber\\
 & &
  +\frac{1}{2}m_{\rho}^2\rho^2
  +3 \Lambda_{\rm{v}}\left(g^2_{\omega}\omega^2\right)
     \left(g^2_{\rho}\rho^2\right),
\end{eqnarray}
and
\begin{eqnarray}
\label{eq:SRMF}
s &=& -\displaystyle{\sum_{i=p,n}\frac{1}{\pi^{2}}
  \int_{0}^{\infty}dk\,k^{2}}
  \left[ f_{i+}^{k}\ln f_{i+}^{k}+\left( 1-f_{i+}^{k}\right)
  \ln \left(1-f_{i+}^{k}\right) \right.  \nonumber \\
& & \left. +f_{i-}^{k}\ln f_{i-}^{k}
  +\left( 1-f_{i-}^{k}\right) \ln \left( 1-f_{i-}^{k}\right) \right].
\end{eqnarray}
Here $M^{\ast}=M+g_{\sigma}\sigma$ is the effective nucleon mass.
$f_{i+}^{k}$ and $f_{i-}^{k}$ ($i=p,n$) denote, respectively, the occupation
probabilities of nucleon and antinucleon at momentum $k$,
which are given by the Fermi-Dirac distribution,
\begin{eqnarray}
\label{eq:firmf}
f_{i\pm}^{k}=\left\{1+\exp \left[ \left( \sqrt{k^{2}+{M^{\ast}}^2}
  \mp \nu_{i}\right)/T\right]
 \right\}^{-1},
\end{eqnarray}
with the kinetic part of the chemical potential $\nu_i$ related to
the chemical potential $\mu_i$ as
\begin{eqnarray}
\nu_{i} = \mu_{i} - g_{\omega}\omega - \frac{g_{\rho}}{2}\tau_{3i}\rho .
\end{eqnarray}
The number density of protons ($i=p$) or neutrons ($i=n$) is calculated by
\begin{equation}
\label{eq:nirmf}
 n_{i}=\frac{1}{\pi^2}
       \int_0^{\infty} dk\,k^2\,\left(f_{i+}^{k}-f_{i-}^{k}\right).
\end{equation}
Using the results of the TM1e model as input in the Thomas--Fermi calculation,
we compute the average free energy density of nonuniform matter, and compare it with
the one of uniform matter. At a given temperature $T$, proton fraction $Y_p$,
and baryon mass density $\rho_B$, the thermodynamically stable state is the
one having the lowest free energy density. We determine the stable state and
the phase transition between nonuniform matter and uniform matter by
minimizing the free energy density.

%%%%%%%%%%%%%%%
\begin{table*}[tbp]
\caption{Coupling constants of the TM1e and TM1 models.}
\begin{center}
\begin{tabular}{lcccccccccccc}
\hline\hline
Model & $g_\sigma$  & $g_\omega$ & $g_\rho$ & $g_{2}$ [fm$^{-1}$] & $g_{3}$ & $c_{3}$ & $\Lambda_{\textrm{v}}$ \\
\hline
TM1e  & 10.0289     & 12.6139    & 13.9714  & $-$7.2325        &0.6183   & 71.3075 & 0.0429  \\
TM1   & 10.0289     & 12.6139    &  9.2644  & $-$7.2325        &0.6183   & 71.3075 & 0.0000  \\
\hline\hline
\end{tabular}
\label{tab:1}
\end{center}
\end{table*}
%%%%%%%%%%%%%%%
%%%%%%%%%%%%%%%
%%%%%%%%%%%%%%%

%%%%%%%%%%%%%%%
\subsection{Thomas--Fermi approximation}
\label{sec:2.2}

At low temperature and subnuclear density region, heavy nuclei are
formed in order to lower the free energy of the system.
For the description of nonuniform matter, we employ the Thomas--Fermi
approximation with a parameterized nucleon distribution,
which was developed by~\citet{oyam93} and used in our previous works~\citep{shen98b,shen11}.
The nonuniform matter is modeled as a mixture of a single species of
heavy nuclei, alpha particles, and free nucleons outside nuclei,
while the leptons are approximated as an ideal relativistic gas separately.
The spherical nuclei are arranged in a body-centered-cubic (BCC) lattice
to minimize the Coulomb lattice energy~\citep{oyam93},
while the Wigner--Seitz cell is introduced to simplify the calculation of free energy.
It is likely that nonspherical nuclei, known as pasta phases, may appear as the
density approaches the phase transition to uniform matter~\citep{avan10,pais12,okam13,bao15}.
The appearance of pasta phases can smooth the transition to uniform
matter (see, e.g.,~\citet{furu13}),
but the effects on thermodynamic quantities in the EOS table
are rather small. For simplicity, we consider only spherical configuration
in constructing the EOS table.

In the Wigner--Seitz cell, a spherical heavy nucleus is located at the center,
while free nucleons and alpha particles exist outside the nucleus.
Each cell is assumed to be charge neutral and the background electron gas is uniform.
The density distribution of particle $i$ ($i=p$, $n$, or $\alpha$) in the cell
is assumed to have the form
\begin{equation}
\label{eq:nitf}
n_i\left(r\right)=\left\{
\begin{array}{ll}
\left(n_i^{\rm{in}}-n_i^{\rm{out}}\right) \left[1-\left(\frac{r}{R_i}\right)^{t_i}
\right]^3 +n_i^{\rm{out}},  &\hspace{0cm} 0 \leq r \leq R_i, \\
n_i^{\rm{out}},  &\hspace{-1.5cm} R_i \leq r \leq R_{\rm{cell}}, \\
\end{array} \right.
\end{equation}
where $r$ denotes the distance from the center of the cell.
$R_{\rm{cell}}$ is the radius of the cell, which is related to the cell
volume $V_{\rm{cell}}$ and the lattice constant $a$ by
$V_{\rm{cell}} = a^3 = 4 \pi R_{\rm{cell}}^3 / 3 =N_B / n_B $
with $N_B$ and $n_{B}$ being the baryon number per cell and the average
baryon number density, respectively.
The baryon mass density is defined as $\rho_B=m_{u} n_B$ with
$m_{u}=931.494$ MeV being the atomic mass unit.
For nonuniform matter at given temperature $T$, proton fraction $Y_p$,
and baryon mass density $\rho_B$, the thermodynamically stable state is
the one with the lowest free energy density, $f=F_{\rm{cell}}/V_{\rm{cell}}$.
The free energy per cell $F_{\rm{cell}}$ is given by
\begin{equation}
\label{eq:fc}
F_{\rm{cell}}=\left(E_{\rm{bulk}}+E_{\rm{surf}}+E_{\rm{Coul}}\right)- T S_{\rm{cell}},
\end{equation}
where the bulk energy $E_{\rm{bulk}}$ and entropy $S_{\rm{cell}}$ are computed by
performing integrations over the cell. The local energy and entropy densities
can be expressed as the sum of contributions from nucleons and alpha particles.
We use the RMF results of the TM1e model for the contributions of nucleons,
while the alpha particles are treated as an ideal Boltzmann gas.
To describe the dissolution of alpha particles at high densities,
the excluded-volume correction is taken into account as described in~\citet{shen11}.
For performing numerical integrations of $E_{\rm{bulk}}$ and $S_{\rm{cell}}$,
we use the tabulated results of the TM1e model given by Equations~(\ref{eq:ERMF})
and~(\ref{eq:SRMF}) as input in the Thomas--Fermi calculation, and then the
corresponding local densities contributed by nucleons are computed from
the input table using a linear interpolation procedure.
The input table is designed to include 871 grid points for the baryon number
density $n_B$ and 1001 grid points for the proton fraction $Y_p$, so that
the linear interpolation can be used with good accuracy.
As for the contribution of alpha particles, it is calculated within the ideal-gas
approximation, where the alpha-particle binding energy $B_{\alpha}=28.3$ MeV is
taken into account~\citep{latt91,shen11}.
Generally, the number density of alpha particles is rather small, and therefore,
the ideal-gas approximation can provide a reasonable description for alpha particles.

In Equation~(\ref{eq:fc}), $E_{\rm{surf}}$ represents the surface energy due to the inhomogeneity
of nucleon distributions. We use the simple form as
\begin{equation}
\label{eq:es}
E_{\rm{surf}}=\int_{\rm{cell}} F_0 \mid \nabla \left( \, n_n\left(r\right)+
    n_p\left(r\right) \, \right) \mid^2 d^3r,
\end{equation}
where the parameter $F_0=70 \, \rm{MeV\,fm^5}$ is the same as that adopted in Shen EOS
with the original TM1 model, which was determined in~\citet{shen98a} by performing
the Thomas--Fermi calculation for finite nuclei so as to reproduce
the gross properties of nuclear masses and charge radii, as described in the Appendix
of~\citet{oyam93}. The reason why we use the same value of $F_0$
in the new EOS4 is because the TM1e model can predict very similar properties of
finite nuclei as the original TM1 model (see Table 2 below), and hence the Thomas--Fermi
calculation in the TM1e model with $F_0=70 \, \rm{MeV\,fm^5}$ is able to reproduce
similar gross properties of nuclear masses and charge radii.
The Coulomb energy per cell $E_{\rm{Coul}}$ is given by
\begin{equation}
\label{eq:ec}
E_{\rm{Coul}}=\frac{1}{2}\int_{\rm{cell}} e \left[n_p\left(r\right)
+2n_{\alpha}\left(r\right)-n_e\right]\,\phi(r) d^3r
\,+\,\triangle E_C,
\end{equation}
where $\phi(r)$ denotes the electrostatic potential calculated
in the Wigner--Seitz approximation and $\triangle E_C$ is the correction term
for the BCC lattice~\citep{oyam93,shen11}.

At given temperature $T$, proton fraction $Y_p$, and baryon mass density $\rho_B$,
we perform the minimization of the free energy density with respect to
independent variables in the parameterized Thomas--Fermi approximation.
To avoid the presence of too many parameters in the minimization procedure,
we use the same parameters $R_p$ and $t_p$ for both proton and
alpha-particle distribution functions. Furthermore, $n_{\alpha}^{\rm{in}}=0$
is adopted, so that alpha particles disappear at the center of the nucleus.
In principle, the nucleon distribution in the Wigner--Seitz cell can be
determined in a self-consistently Thomas--Fermi approximation,
where the set of coupled equations is solved by the iteration method
in coordinate space~\citep{zhang14}.
However, the self-consistent Thomas--Fermi calculation
requires much more computational effort than the parameterized Thomas--Fermi
approximation. In our previous work~\citep{zhang14}, we made a detailed
comparison between the self-consistent Thomas--Fermi approximation
and the parameterized Thomas--Fermi approximation, which showed that
the differences in thermodynamic quantities between these two methods
are negligible and would not affect the general behavior of the EOS.
Therefore, we prefer to employ the parameterized Thomas--Fermi approximation
in the present calculation. Furthermore, it is also helpful for examining
the effects of symmetry energy by comparing EOS4 with EOS2 based on
the same method.

After the thermodynamically favorable state is determined in the minimization procedure,
we calculate the thermodynamic quantities like the pressure and chemical potentials
from the free energy per baryon $F \left(T,Y_{p},n_{b}\right)$ over the full range
of the EOS table by the thermodynamic relations:
\begin{eqnarray}
 p\left(T,Y_{p},n_{B}\right) &=& \left[ n_B^{2} \frac{\partial F} {\partial n_B} \right]_{T,Y_{p}},
\label{eq:ppre} \\
\mu_{p}\left(T,Y_{p},n_{B}\right) &=& \left[\frac{\partial \left( n_{B} F\right)}{\partial n_p} \right]_{T,n_n},
\label{eq:pmup} \\
\mu_{n}\left(T,Y_{p},n_{B}\right) &=& \left[\frac{\partial \left( n_{B} F\right)}{\partial n_n} \right]_{T,n_p},
\label{eq:pmun}
\end{eqnarray}
where $n_{p}=Y_{p}n_{B}$ and $n_{n}=\left(1-Y_{p}\right)n_{B}$ are the average number densities
of protons and neutrons, respectively. The final EOS table contains not only thermodynamic
quantities but also compositions of matter and other information.
For convenience in use and comparison, the new EOS4 is designed
to have the same tabular form as EOS2, while the definitions of the physical
quantities in the EOS table have been given in Appendix A of~\citet{shen11}.

Compared to the treatment of nonuniform matter in~\citet{shen11},
the results of the TM1e model are used as input in the Thomas--Fermi calculation
of EOS4, instead of the original TM1 model used in EOS2.
The different density dependence of symmetry energy between TM1e ($L=40$ MeV)
and TM1 ($L=110.8$ MeV) would lead to more significant effects
in low $Y_p$ region.
It is interesting to make a detailed comparison between EOS4 and EOS2,
so that we can explore the influences of symmetry energy and
its density dependence on properties of the EOS for astrophysical simulations.

%%%%%%%%%%%%%%%%%%%%%%%%%%%%%%%%%%%%%%%%%%%%%%%%%%%%%%%%%%%%%%%%%%%%%%%%%%%%%%%%
\section{Results}
\label{sec:3}

We construct the new EOS4 based on the TM1e model with $L=40$ MeV covering a wide
range of temperature $T$, proton fraction $Y_p$, and baryon mass density $\rho_B$
for numerical simulations of core-collapse supernovae and neutron-star mergers.
For convenience in practical use, we provide the new EOS4 in the same tabular form
within the ranges as given in Table 1 of~\citet{shen11}.
All physical quantities included in the EOS table have been defined in
Appendix A of~\citet{shen11}.
Compared to the EOS2 based on the original TM1 model in~\citet{shen11},
the new EOS4 is more compatible with both experimental nuclear data and recent
observations of neutron stars.
In Table~\ref{tab:2}, we present some properties of nuclear symmetry energy,
finite nuclei, and neutron stars, so as to examine the compatibility of the models
with current constraints and experimental data.
It is shown that the results of finite nuclei in the TM1e and TM1 models are very
similar to each other and in good agreement with experimental data.
On the other hand, the TM1e model provides much smaller radius and tidal
deformability for a $1.4 M_\odot$ neutron star, which is more consistent
with the current constraints.
It is reasonable that different behaviors of the symmetry energy between these
two models have more pronounced effects for neutron-rich objects like neutron stars.
More detailed properties of neutron stars obtained in the TM1e model have been
reported in our recent study~\citep{ji19}.
%%%%%%%%%%%%%%%
\begin{table*}[htb]
\caption{Properties of symmetry energy, finite nuclei, and neutron stars
in the TM1e and TM1 models.
$E_{\rm{sym}}$ and $L$ are the nuclear symmetry energy and its slope
parameter at saturation density, respectively.
The binding energy per nucleon $E/A$, charge radius $r_{\text{c}}$,
and neutron-skin thickness $\triangle r_{\text{np}}$ of $^{208}$Pb
obtained in the RMF approach and Thomas--Fermi (TF) approximation
are compared with the experimental data in the last column.
$M_{\mathrm{max}}$ is the maximum mass of neutron stars,
while $R_{1.4}$ and $\Lambda_{1.4}$ denote the radius and tidal
deformability for a $1.4 M_\odot$ neutron star, respectively.}
\begin{center}
\begin{tabular}{llccl}
\hline\hline
                &                      & EOS4(TM1e) & EOS2(TM1) &  constraints\\
\hline
symmetry energy & $E_{\rm{sym}}$ [MeV] & 31.38       & 36.89 & $31.7\pm 3.2$~\citep{oert17}   \\
                & $L$ [MeV]            & 40          & 110.8 & $58.7\pm 28.1$~\citep{oert17}  \\
\hline
finite nuclei (RMF)  & $E/A$ ($^{208}$Pb) [MeV]             & 7.88 & 7.88 & 7.87~\citep{audi03}  \\
               & $r_{\rm{c}}$ ($^{208}$Pb) [fm]             & 5.56 & 5.54 & 5.50~\citep{ange13}  \\
               & $\triangle r_{\rm{np}}$ ($^{208}$Pb) [fm]  & 0.16 & 0.27 & $0.33^{+0.16}_{-0.18}$~\citep{abra12}
\vspace{0.2cm}\\
finite nuclei (TF)  & $E/A$ ($^{208}$Pb) [MeV]              & 8.05 & 8.08 & 7.87~\citep{audi03}  \\
               & $r_{\rm{c}}$ ($^{208}$Pb) [fm]             & 5.68 & 5.65 & 5.50~\citep{ange13}  \\
               & $\triangle r_{\rm{np}}$ ($^{208}$Pb) [fm]  & 0.10 & 0.21 & $0.33^{+0.16}_{-0.18}$~\citep{abra12} \\
\hline
neutron stars  & $M_{\rm{max}}$ [$M_\odot$] & 2.12  &  2.18 &  $1.928 \pm 0.017$~\citep{fons16} \vspace{-0.0cm}\\
               &                            &       &       &  $2.01  \pm 0.04$~\citep{anto13} \vspace{-0.0cm}\\
               &                            &       &       &  $2.14^{+0.10}_{-0.09}$~\citep{crom19} \\
               & $R_{1.4}$ [km]           & 13.1  &  14.2 &    $10.5<R_{1.4}<13.3$~\citep{abbo18} \\
               &                            &       &       &  $12.0<R_{1.4}<13.45$~\citep{most18} \\
               & $\Lambda_{1.4}$          & 652   &  1047 &   $<800$~\citep{abbo17} \\
               &                            &       &       &  $190^{+390}_{-120}$~\citep{abbo18} \\\hline\hline
\end{tabular}
\label{tab:2}
\end{center}
\end{table*}

%%%%%%%%%%%%%%%
\begin{figure}[htbp]
\centering
\includegraphics[width=8.6 cm]{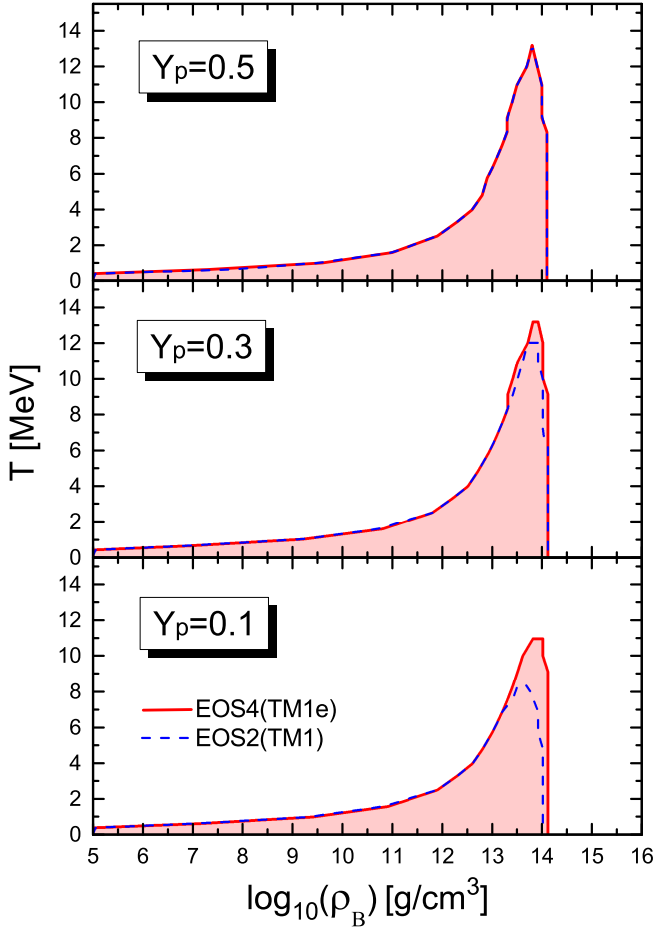}
\caption{Phase diagram in the $\rho_B$--$T$ plane for $Y_p=0.1$, $0.3$, and $0.5$.
The shaded region corresponds to the nonuniform matter phase where heavy
nuclei are formed. }
\label{fig:3TRho}
\end{figure}
%%%%%%%%%%%%%%%
\begin{figure}[htbp]
\centering
\includegraphics[width=8.6 cm]{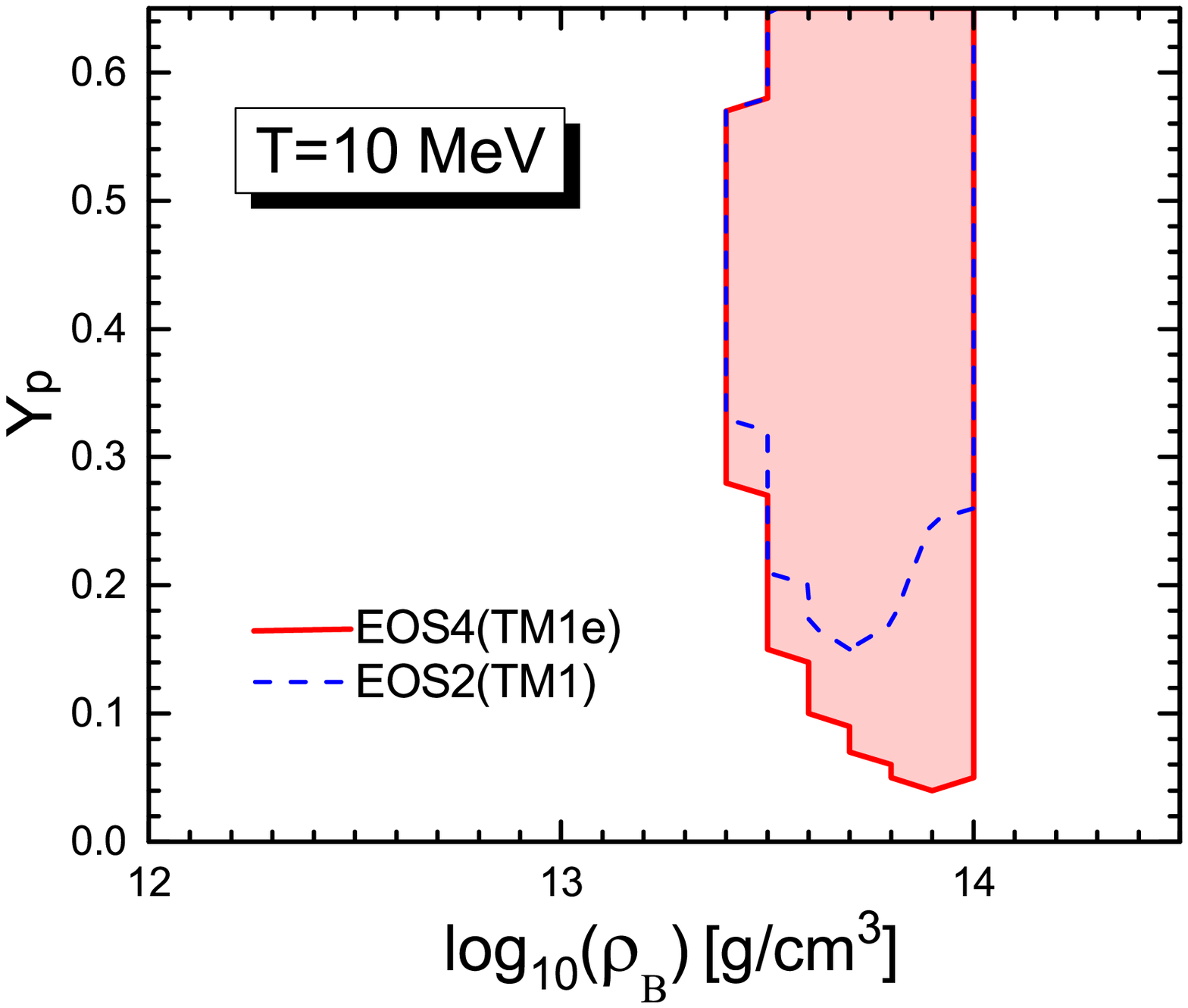}
\caption{Phase diagram in the $\rho_B$--$Y_p$ plane at $T=10$ MeV.
The shaded region corresponds to the nonuniform matter phase where heavy
nuclei are formed. }
\label{fig:4YpRho}
\end{figure}
%%%%%%%%%%%%%%%

To build the EOS table for astrophysical simulations, we perform the free energy
minimization at each $T$, $Y_p$, and $\rho_B$ for both nonuniform matter and
uniform matter. The thermodynamically favorable state is the one with the
lowest free energy density among all configurations considered.
The phase transition is determined by comparing the free energy density
between nonuniform matter and uniform matter.
In Figure~\ref{fig:3TRho}, we show the phase diagram in the $\rho_B$--$T$
plane for $Y_p=0.1$, $0.3$, and $0.5$ obtained in EOS4 (red solid lines)
which is compared with that of EOS2 (blue dashed lines).
One can see that the nonuniform matter phase with heavy nuclei can exist
only at low temperature and subnuclear density region.
At low densities, the uniform matter consists of a free nucleon gas together
with a small fraction of alpha particles. As the density increases,
heavy nuclei are formed in the nonuniform matter phase to lower
the free energy. When the density is beyond $\sim 10^{14.1}\,\rm{g/cm^{3}}$,
heavy nuclei dissolve and the favorable state becomes the uniform nuclear matter.
The density range of the nonuniform matter phase depends on both $T$ and $Y_p$.
As the temperature increases, the onset density of nonuniform matter increases
significantly, while the transition from nonuniform matter to uniform matter
is almost independent of $T$. When the temperature reaches the critical value $T_c$,
the nonuniform matter phase disappears completely, i.e. heavy nuclei cannot be
formed at $T>T_c$.

It is interesting to note the effects of symmetry energy
on the boundary of nonuniform matter.
For the case of $Y_p=0.5$ shown in the top panel of Figure~\ref{fig:3TRho},
there is no visible difference between EOS4(TM1e) and
EOS2(TM1) due to the same isoscalar properties in the two models.
For the case of $Y_p=0.1$ shown in the bottom panel,
the critical temperature $T_c$ in EOS4(TM1e) is
significantly higher than the one obtained in EOS2(TM1).
Furthermore, the transition density to uniform matter in EOS4(TM1e) is slightly
larger than that in EOS2(TM1). This is consistent with the correlation between
the symmetry energy slope and the crust-core transition density of neutron
stars~\citep{bao15}.
In Figure~\ref{fig:4YpRho}, we show the density range of nonuniform matter
as a function of $Y_p$ at $T=10$ MeV.
It is seen that there is clear difference between EOS4(TM1e) and EOS2(TM1)
in low $Y_p$ region, where the behavior of symmetry energy plays an important role
in determining the properties of neutron-rich matter. One can see that heavy nuclei
do not appear in EOS2(TM1) at $T=10$ MeV for $Y_p<0.15$, whereas the
nonuniform matter phase exists until $Y_p\sim 0.04$ in EOS4(TM1e).
Similar effects of the symmetry energy and its slope on the phase diagram were also
observed in~\citet{toga17}, where the authors constructed the EOS table using
a non-relativistic variational method based on realistic nuclear forces.
It is interesting to find this similarity for both non-relativistic and relativistic
many-body frameworks with small $L$ values.

%%%%%%%%%%%%%%%
\begin{figure}[htbp]
\centering
\includegraphics[width=8.6 cm]{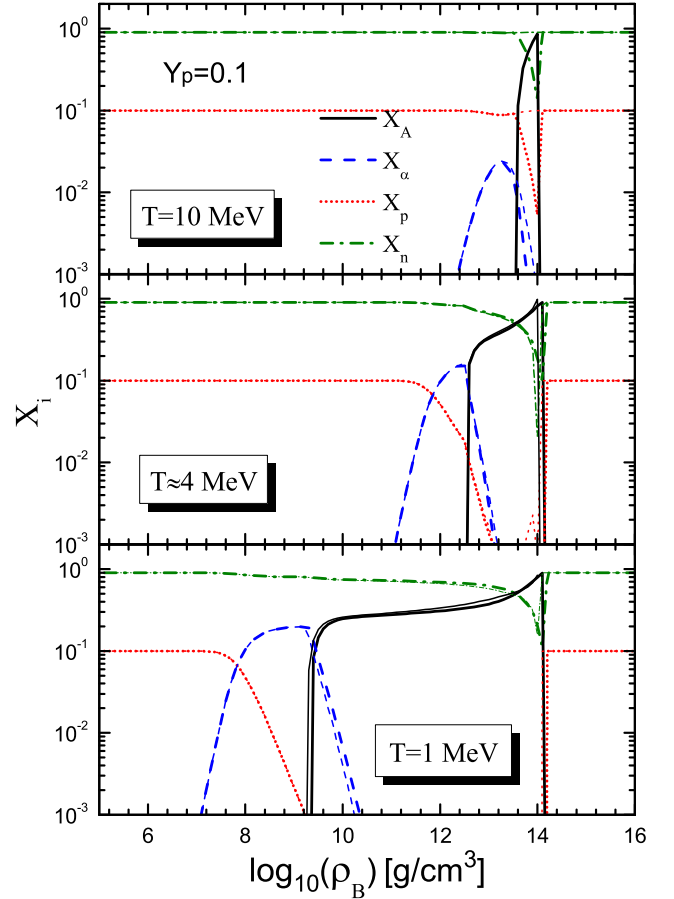}
\caption{Fraction of neutrons (green dash-dotted line), protons (red dotted line), alpha
particles (blue dashed line), and heavy nuclei (black solid line) as a function of the
baryon mass density $\rho_B$ for $Y_p=0.1$ at $T=1$, $4$, and $10$ MeV.
The results obtained in EOS4(TM1e) and EOS2(TM1) are shown by thick and thin
lines, respectively. }
\label{fig:5XiRho}
\end{figure}
%%%%%%%%%%%%%%%

In Figure~\ref{fig:5XiRho}, we show the fractions of neutrons, protons, alpha particles,
and heavy nuclei as a function of the baryon mass density $\rho_B$ for $Y_p=0.1$
at $T=1$, $4$, and $10$ MeV. At low densities, the matter is a uniform gas of
neutrons and protons together with a small fraction of alpha particles.
The alpha-particle fraction $X_{\alpha}$ increases with increasing $\rho_B$
before the formation of heavy nuclei, but it rapidly decreases in the nonuniform
matter where heavy nuclei use up most of the nucleons.
When the density increases beyond $\sim 10^{14.1}\,\rm{g/cm^{3}}$,
heavy nuclei dissolve and the matter is composed of uniform neutrons and protons.
In general, the results of EOS2 (thin lines) are just slightly different from those
of EOS4 (thick lines). In the case of $T=10$ MeV (top panel), heavy nuclei
do not appear in EOS2 with the TM1 model, but alpha particles exist at intermediate densities.
This is different from the results of EOS4, where heavy nuclei are formed in the
density range $10^{13.6}\leq \rho_B \leq 10^{14.0}\,\rm{g/cm^{3}}$ with the TM1e
model. Due to the formation of heavy nuclei, $X_{n}$ and $X_{p}$ in this density
range are much different between EOS4 and EOS2.

%%%%%%%%%%%%%%
\begin{figure}[htb]
\centering
\includegraphics[width=8.6 cm]{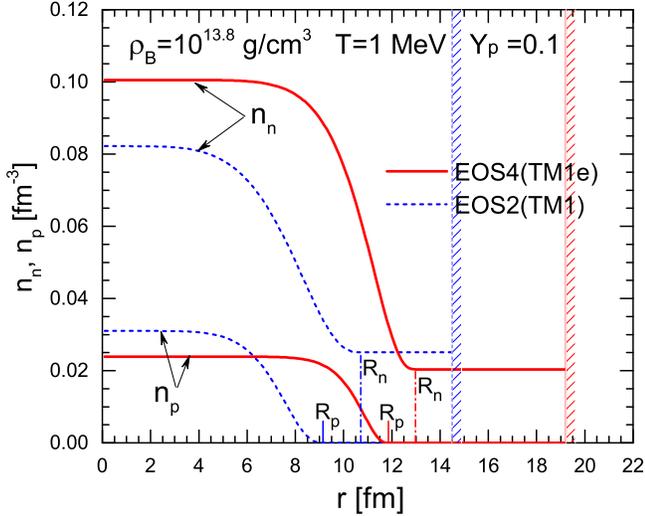}
\caption{Density distributions of protons and neutrons inside the
Wigner--Seitz cell for the case of $T=1$ MeV and $Y_p=0.1$
at $\rho_B = 10^{13.8}\, \rm{g/cm^{3}}$.
The results obtained in EOS4 (red solid lines)
are compared with those of EOS2 (blue dashed lines).
The radius of the Wigner--Seitz cell is indicated by the hatch,
while the radius of the heavy nucleus is shown by the dash-dotted line. }
\label{fig:6dis}
\end{figure}

%%%%%%%%%%%%%%
\begin{figure}[htb]
\centering
\includegraphics[width=8.6 cm]{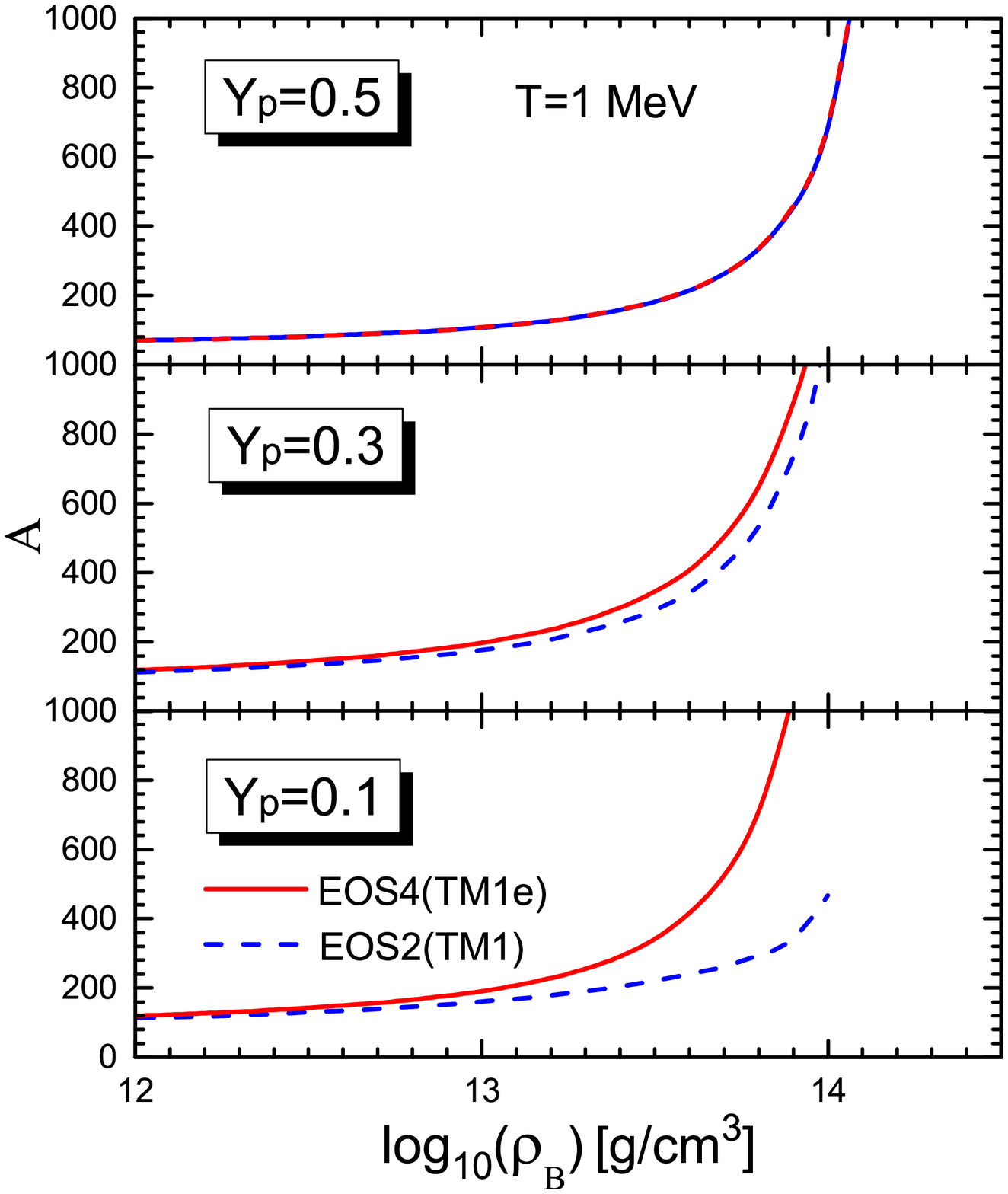}
\caption{Nuclear mass number $A$ as a function of the baryon mass density $\rho_B$
at $T=1$ MeV for $Y_p=0.5$, $0.3$, and $0.1$. The results obtained in EOS4 (red solid lines)
are compared with those of EOS2 (blue dashed lines). }
\label{fig:7A}
\end{figure}

\begin{figure}[htb]
\centering
\includegraphics[width=8.6 cm]{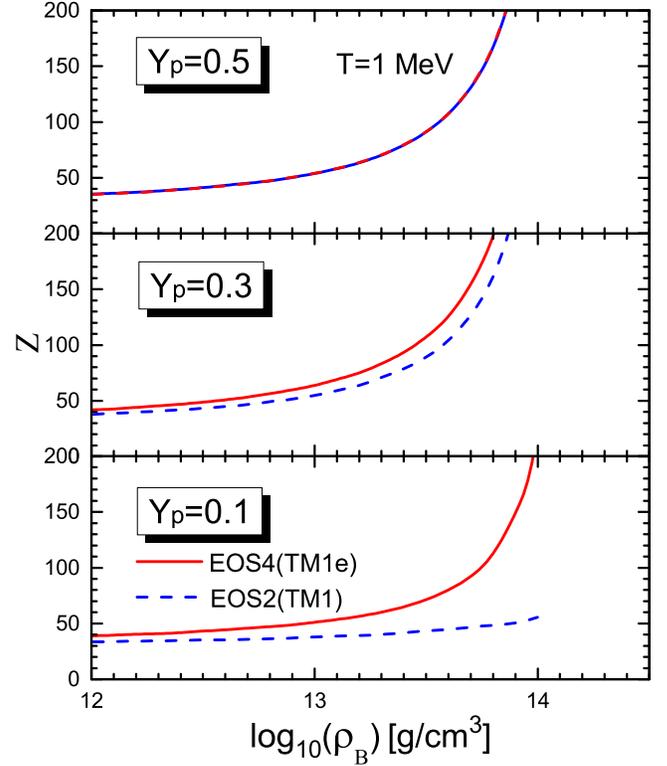}
\caption{Charge number $Z$ as a function of the baryon mass density $\rho_B$
at $T=1$ MeV for $Y_p=0.5$, $0.3$, and $0.1$. The results obtained in EOS4 (red solid lines)
are compared with those of EOS2 (blue dashed lines). }
\label{fig:8Z}
\end{figure}
%%%%%%%%%%%%%%%

In nonuniform matter, the properties of heavy nuclei are determined by
minimizing the free energy density in the parameterized Thomas--Fermi approximation.
We display in Figure~\ref{fig:6dis} the resulting density distributions of protons
and neutrons inside the Wigner--Seitz cell for the case of $T=1$ MeV and $Y_p=0.1$
at $\rho_B = 10^{13.8}\, \rm{g/cm^{3}}$.
The radius of the Wigner--Seitz cell is indicated by the hatch,
while the radius of the heavy nucleus is shown by the dash-dotted line.
The results obtained in EOS4 (red solid lines)
are compared with those of EOS2 (blue dashed lines).
It is shown that both the cell radius $R_{\rm{cell}}$ and the neutron
radius $R_n$ (i.e., the radius of the heavy nucleus) obtained in EOS4
are larger than those in EOS2. Furthermore, the neutron-skin
thickness, $R_n-R_p$, is relatively small in the case of EOS4.
This is because the TM1e model used in EOS4 has a smaller symmetry energy
slope $L=40$ MeV than the value of $L=110.8$ MeV in the TM1 model of EOS2.
It is well known that the neutron-skin thickness of finite nuclei is
positively correlated to the symmetry energy slope $L$.
On the other hand, the density distributions, $n_n$ and $n_p$, are also
largely affected by the symmetry energy slope $L$.
The dripped neutron density $n_n^{\rm{out}}$ of EOS4 is smaller than
that of EOS2, while the neutron density at the center $n_n^{\rm{in}}$
is much larger. This tendency can be understood from different behaviors
of the symmetry energy between TM1e and TM1 models.
As shown in Figure~\ref{fig:2Esym}, the TM1e model has larger $E_{\text{sym}}$
at low densities but smaller $E_{\text{sym}}$ at high densities
compared to the TM1 model. Therefore, the TM1e model results in
relatively larger $n_n^{\rm{in}}$ and smaller $n_n^{\rm{out}}$ than
the TM1 model. It is seen that the density gradient in EOS4 is larger
than that in EOS2, which leads to larger surface energy and nuclear radius.
A similar behavior was also reported in~\citet{toga17},
where the authors used the model with small $L$ based on realistic
nuclear forces and compared to the results of EOS2.
In Figures~\ref{fig:7A} and \ref{fig:8Z}, we show respectively the nuclear mass
number $A$ and charge number $Z$ as a function of the baryon mass density $\rho_B$
at $T=1$ MeV for $Y_p=0.5$, $0.3$, and $0.1$.
It is seen that both $A$ and $Z$ weakly depend on $\rho_B$ at low densities and
rapidly increase before the transition to uniform matter.
There are significant differences between EOS4 and EOS2 for small $Y_p$.
The values of $A$ and $Z$ obtained in EOS4 are larger than those of EOS2.
This is because the TM1e model with a small $L$ results in a large nuclear radius
as shown in Figure~\ref{fig:6dis}, which implies more protons
and neutrons are bound inside the heavy nucleus.
The differences of heavy nuclei between EOS4 and EOS2 may affect the neutrino
transport and emission in core-collapse supernovae, which need to be explored
in further studies.

%%%%%%%%%%%%%%
\begin{figure}[htb]
\centering
\includegraphics[width=8.6 cm,clip]{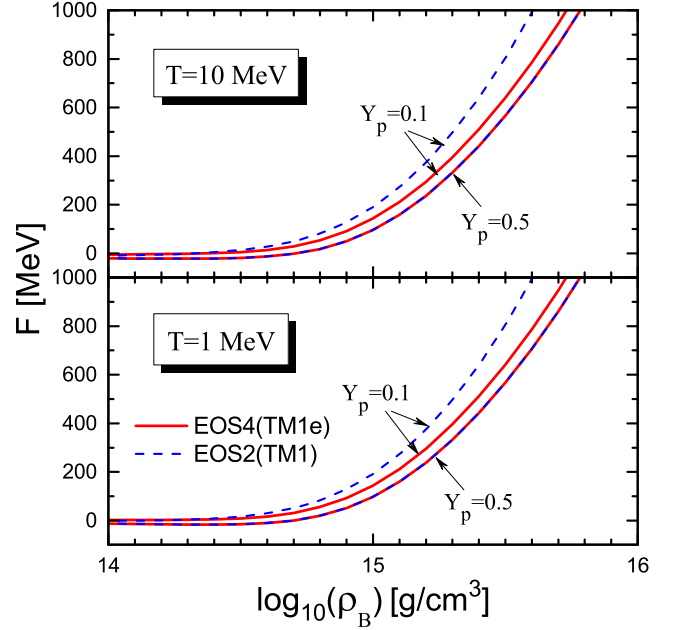}
\caption{Free energy per baryon $F$ as a function of the baryon mass
density $\rho_B$ with $Y_p=0.1$ and $0.5$ at $T=1$ and $10$ MeV.
The results obtained in EOS4 (red solid lines)
are compared with those of EOS2 (blue dashed lines). }
\label{fig:9F}
\end{figure}

%%%%%%%%%%%%%%
\begin{figure}[htb]
\centering
\includegraphics[width=8.6 cm,clip]{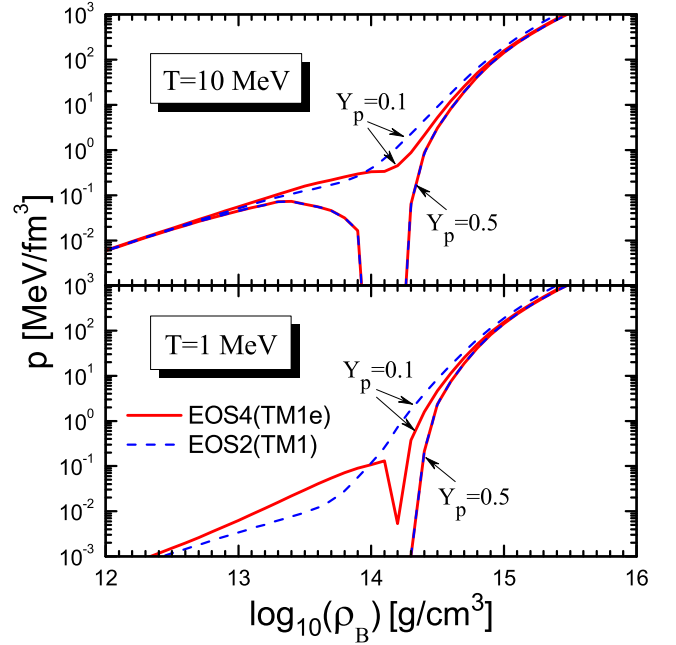}
\caption{Same as Figure~\ref{fig:9F}, but for the pressure $p$. }
\label{fig:10p}
\end{figure}

%%%%%%%%%%%%%%
\begin{figure}[htb]
\centering
\includegraphics[width=8.6 cm,clip]{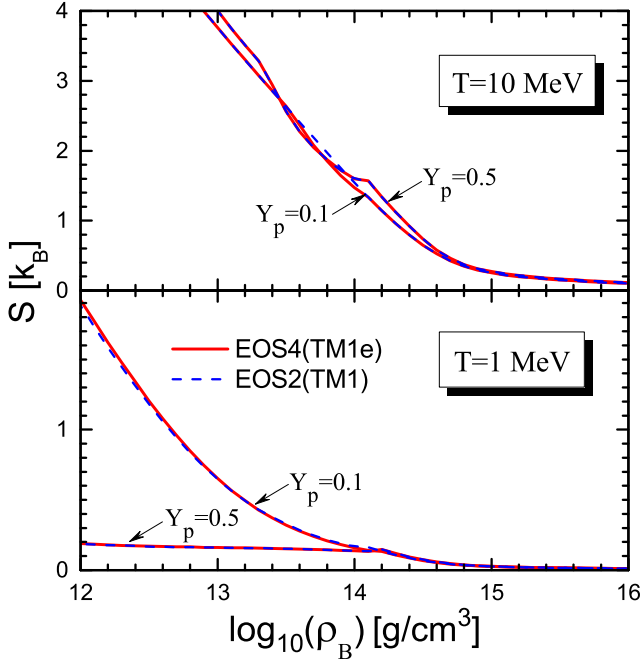}
\caption{Same as Figure~\ref{fig:9F}, but for the entropy per baryon $S$. }
\label{fig:11S}
\end{figure}

It is essential to discuss the effects of symmetry energy on the thermodynamic
quantities in the EOS table.
In Figure~\ref{fig:9F}, we show the free energy per baryon $F$ as a function of the
baryon mass density $\rho_B$ for $Y_p=0.1$ and $0.5$ at $T=1$ and $10$ MeV.
The results in EOS4(TM1e) are shown by solid lines,
while those in EOS2(TM1) are displayed by dashed lines for comparison.
There is almost no difference between EOS4 and EOS2 for the case of $Y_p=0.5$
due to the same isoscalar properties of the two models.
In the case of $Y_p=0.1$, the values of $F$ in EOS4 are smaller than those in EOS2,
and their difference increases with increasing $\rho_B$.
This is because the TM1e model has smaller symmetry energy than the TM1 model
at high densities, which leads to smaller free energy in neutron-rich matter.
Comparing the cases between $T=1$ MeV and $T=10$ MeV, the tendencies of the free
energy are very similar to each other. This implies that the dependence of
the symmetry energy effect on $T$ is rather weak.
We plot in Figure~\ref{fig:10p} the pressure $p$ as a function of $\rho_B$
for $Y_p=0.1$ and $0.5$ at $T=1$ and $10$ MeV.
The pressure is calculated from the derivative of the free energy, as given
in Equation~(\ref{eq:ppre}). Due to the formation of heavy nuclei
in nonuniform matter, the pressure has a rapid drop as shown in the case
of $Y_p=0.5$ in the top panel (Note that this drop does not appear when contributions
from leptons and photons are added).
In contrast, the pressure for $Y_p=0.1$ is
much smooth due to less fraction of heavy nuclei.
It is noticed that there is a clear discontinuity in EOS4(TM1e) around the
phase transition to uniform matter $\sim 10^{14.2}\,\rm{g/cm^{3}}$ for the case
of $Y_p=0.1$ and $T=1$ MeV. In fact, this discontinuity is also found
in other cases (see, e.g., Figure 7 of~\citet{shen98b} and Figure 14 of~\citet{toga17}).
This is because the phase transition is determined by minimizing the free energy,
and as a result, the free energy shown in Figure~\ref{fig:9F} is a smooth function
of the density. However, the pressure calculated from the first derivative of
the free energy may exhibit a discontinuity at the first-order phase transition~\citep{pais14}.
Compared to the results of EOS2 shown by dashed lines, the pressure of uniform matter
beyond $\sim 10^{14.1}\,\rm{g/cm^{3}}$ in EOS4 for $Y_p=0.1$ is relatively small,
so the discontinuity is more obvious in this case.
Generally, the pressure at high densities obtained in EOS4 is lower than that in EOS2,
which is a result of small density dependence of symmetry energy in the TM1e model.
Therefore, the new EOS4 is softer than EOS2 due to different behaviors
of the symmetry energy between these two models.
In Figure~\ref{fig:11S}, we show the entropy per baryon $S$ as a function of
$\rho_B$ for $Y_p=0.1$ and $0.5$ at $T=1$ and $10$ MeV.
At $T=1$ MeV, the values of $S$ for $Y_p=0.5$ are much smaller than those
for $Y_p=0.1$. This is because most of nucleons exist inside heavy nuclei for
the case of $Y_p=0.5$, while there is a large fraction of free neutrons
for $Y_p=0.1$ as shown in Figure~\ref{fig:5XiRho}.
At $T=10$ MeV, the difference of $S$ between $Y_p=0.5$ and $Y_p=0.1$
is relatively small, because the formation of heavy nuclei becomes less
important as the temperature increases.
It is found that the difference of symmetry energy between TM1e and TM1 models
has minor influence on the entropy, and as a result, the behavior of $S$
obtained in EOS4 is very similar to that in EOS2.
Generally, the TM1e model with $L=40$ MeV leads to visible differences in EOS4
from EOS2 for $Y_p\leq 0.3$, and the difference increases as the matter
becomes more neutron-rich.

%%%%%%%%%%%%%%%%%%%%%%%%%%%%%%%%%%%%%%%%%%%%%%%%%%%%%%%%%%%%%%%%%%%%%%%%%%%%%%%%
\section{Summary}
\label{sec:4}

In this work, we constructed a new EOS table (EOS4) based on an extended
TM1 model with $L=40$ MeV (referred to as the TM1e model) for astrophysical
simulations of core-collapse supernovae and neutron-star mergers.
Following the method described in our previous study~\citep{shen11},
we employed the Thomas--Fermi approximation with a parameterized nucleon
distribution for the description of nonuniform matter,
which is modeled as a mixture of a single species of heavy nuclei,
alpha particles, and free nucleons outside nuclei.
At given temperature $T$, proton fraction $Y_p$, and baryon mass
density $\rho_B$, we perform the minimization of the free energy density
with respect to independent variables involved, so as to determine
the thermodynamically stable state with the lowest free energy.
For convenience in use and comparison, the new EOS4 was designed in the
same tabular form as the previous version EOS2 presented in~\citet{shen11}.
Now, both EOS4 and EOS2 are available at
{\it http://my.nankai.edu.cn/wlxy/sh\_en/list.htm},
{\it http://user.numazu-ct.ac.jp/$^{\sim}$sumi/eos/index.html},
and {\it http://doi.org/10.5281/zenodo.3612487}.

The main difference between the new EOS4 in this work and the previous EOS2
in~\citet{shen11} is that the TM1e model with a small
slope parameter $L=40$ MeV was used in EOS4 instead of the original
TM1 model with $L=110.8$ MeV adopted in EOS2.
The different behaviors of the symmetry energy between TM1e
and TM1 lead to visible impacts on various aspects of the EOS for
astrophysical simulations, especially in the neutron-rich region.
The effects of the symmetry energy and its slope observed in this work
are consistent with those reported in~\citet{toga17}.

The present work was motivated by recent developments in astrophysical observations,
such as the binary neutron-star merger GW170817, which provided new constraints
on the tidal deformability and radii of neutron stars.
It is likely that the TM1 model with $L=110.8$ MeV used in EOS2 predicts too large
neutron-star radii compared to the current observations. Therefore,
we prefer to revise our EOS table by employing the TM1e model with
$L=40$ MeV, which could provide much smaller neutron-star radius.
It is well known that the neutron-star radius is positively correlated to the
symmetry energy slope $L$. By introducing an additional $\omega$-$\rho$ coupling
term, it is possible to modify the density dependence of the symmetry energy
according to the constraints from astrophysical observations and terrestrial
nuclear experiments.
In the TM1e model, we adjusted simultaneously two parameters associated to
the $\rho$ meson, and as a result, the slope parameter $L=40$ MeV and the
symmetry energy $E_{\text{sym}}=31.38$ MeV at saturation density were achieved,
which fall well within the constraints from various observations.
It is noteworthy that the TM1e model provides the same properties of symmetric
nuclear matter and similar binding energies of finite nuclei as the original
TM1 model, whereas the density dependence of the symmetry energy is very
different. This choice allows us to explore
the effect solely from the symmetry energy without interference of
the isoscalar part.

To examine the effect of symmetry energy, we made a detailed comparison
between the new EOS4 and previous EOS2.
It was found that the TM1e model used in EOS4 could predict relatively larger
region of nonuniform matter and softer EOS in the neutron-rich region
compared with the original TM1 model used in EOS2.
In the case of EOS4, the critical temperature, where the nonuniform matter
phase disappears completely, is clearly higher than the one in EOS2
for the case of low $Y_p$.
Furthermore, the transition density to uniform matter in EOS4 is slightly
larger than that in EOS2. In nonuniform matter, the mass number $A$ and
charge number $Z$ of heavy nuclei obtained in EOS4 were found to be
larger than those of EOS2. We also found noticeable differences in the
thermodynamic quantities like the free energy and pressure, especially
for neutron-rich matter at high densities. All these differences between
EOS4 and EOS2 become more significant as $Y_p$ decreases.
This is because the TM1e and TM1 models have the same isoscalar properties
but different density dependence of the symmetry energy.

It is interesting and important to explore the effects of symmetry energy on
astrophysical phenomena such as core-collapse supernovae and neutron-star mergers.
In our recent work~\citep{sumi19}, we have studied the influence of symmetry
energy and its density dependence in numerical simulations of
gravitational collapse of massive stars and cooling of protoneutron stars
by using a hybrid EOS, where the TM1e model was adopted for
uniform matter at densities above $\sim 10^{14}\,\rm{g/cm^{3}}$
combined with the previous EOS2 of nonuniform matter at low densities.
While the TM1e EOS at high densities is shown to have major effects on the birth
of neutron stars in neutron-rich regions, the full table of TM1e EOS including
the part of low densities will have influence on collapse and bounce of supernova
cores, by non-trivial feedback through compositional changes with neutrino reactions,
thereby the outcome of the (non)explosion and the compact object formation.
The numerical simulation of core-collapse supernovae and the analysis of
symmetry energy effects using the new EOS4 are currently underway.

%%%%%%%%%%%%%%%%%%%%%%%%%%%%%%%%%%%%%%%%%%%%%%%%%%%%%%%%%%%%%%%%%%%%%%%%%%%%%%%%
\acknowledgments

We would like to thank H. Toki, K. Oyamatsu, S. Yamada, H. Togashi, S. Furusawa,
and Y. Sekiguchi for fruitful discussions and suggestions on the EOS tables
and their applications.
This work was supported in part by the National Natural Science Foundation of
China (Grants No. 11675083 and No. 11775119).
K.S. is supported by Grant-in-Aid for Scientific Research (15K05093, 19K03837)
and Grant-in-Aid for Scientific Research on Innovative areas
\textquotedblleft Gravitational wave physics and astronomy:Genesis\textquotedblright\
(17H06357, 17H06365)
from the Ministry of Education, Culture, Sports, Science and Technology (MEXT), Japan.
KS also acknowledges the high performance computing resources
at KEK, RCNP, Osaka University and YITP, Kyoto University.

%%%%%%%%%%%%%%%%%%%%%%%%%%%%%%%%%%%%%%%%%%%%%%%%%%%%%%%%%%%%%%%%%%%%%%%%%%%%%%%%

%%%%%%%%%%%%%%%%%%%%%%%%%%%%%%%%%%%%%%%%%%%%%%%%%%%%%%%%%%%%%%%%%%%%%%%%%%%%%%%%
\end{document}